\documentclass{iopconfser}

\usepackage{graphicx}
\usepackage{hyperref}
\usepackage{url}

\begin{document}

\title{Astroaccesible: A multi-messenger outreach for a multi-messenger science}

\author{Enrique P{\'e}rez-Montero$^{1}$}

\affil{$^1$Instituto de Astrof\'{i}sica de Andaluc\'{i}a - CSIC, Apdo. 3004, Granada, Spain}

\email{epm@iaa.es}

\begin{abstract}
This contribution summarizes the main activities and objectives of the outreach project {\em Astroaccesible}, whose main aim is to carry the teaching and diffusion of astronomy among all kinds of collectives, focusing on blind and visually impaired (BVI) people. This project is led by a blind astronomer and aims to use a variety of resources based on different sensory channels, avoiding limiting the transmission of concepts to visual perception. This principle favors inclusion and benefits everyone, as the information is not presented using just one channel. This strategy is especially convenient for the nowadays typical data acquisition, where a variety of sources of information, not solely based on the collection of different spectral domains of electromagnetic radiation, is used. Moreover, the study of new multi-messenger astronomy could be much better understood using a multi-messenger teaching approach, favoring inclusion, motivation, and creativity.
\end{abstract}

\section{Introduction}

The teaching and outreach of science for every public are essential for the formation and establishment of a general culture for any individual. The constant scientific and technological advancements being made nowadays must be accompanied by appropriate dissemination and explanation for the entire society. This fundamental objective, explicitly recognized in several declarations such as the Universal Declaration of Human Rights (1948) and the Convention on the Rights of Persons with Disabilities (2006), cannot be satisfactorily achieved without including all impaired people and those who, due to their characteristics, find barriers to accessing information \cite{ortiz11}.

Astronomy, which includes astrophysics and cosmology, is one of the fastest-growing branches of science and is providing numerous new scientific results. This growth is partly driven by investments in new facilities for observing the Universe, such as large telescopes, space observatories, and missions to objects in our Solar System. The new data and findings from these facilities always have a significant impact on the media.

Given that a large portion of the dissemination of these results is predominantly done through visual media (e.g., pictures, videos, animations, graphics), blind and visually impaired (BVI) people are often excluded from having full and satisfactory access to all this content. Although there is a growing awareness of the need to adapt scientific information for teaching or outreach, we are still far from achieving the necessary objectives to make all content fully accessible for everyone.

Moreover, in the last decades, we have witnessed how the acquisition of observational information coming from the huge heterogeneity of sources and ranges present in the Universe has widened, opening the gate to a new multi-messenger astronomy, which includes data from gravitational waves, cosmic rays, and neutrinos, presenting both a challenge and an opportunity for inclusion. Adapting these new types of data for a wider audience requires innovations in the presentation and teaching of science, fostering inclusion and benefiting everyone by presenting information through multiple sensory channels. This is crucial not only for BVI individuals but also for enhancing understanding and motivation in the study of multi-messenger astronomy.

\section{The inclusive aspect of the project Astroaccesible}

The outreach project {\em Astroaccesible}\footnote{The different activities of the project can be consulted in Spanish in \url{https://astroaccesible.iaa.es}}  \cite{astroaccesible} started as an initiative to teach basic concepts of astronomy among the collective of BVI. Although these first activities had a character mainly restricted to impaired people, the project has incorporated an inclusive aspect \cite{PM19_NatAst} whose benefits have been shared and transmitted to other educators or professionals of outreach.

\begin{figure}[t]
    \centering
    \includegraphics[width=\textwidth]{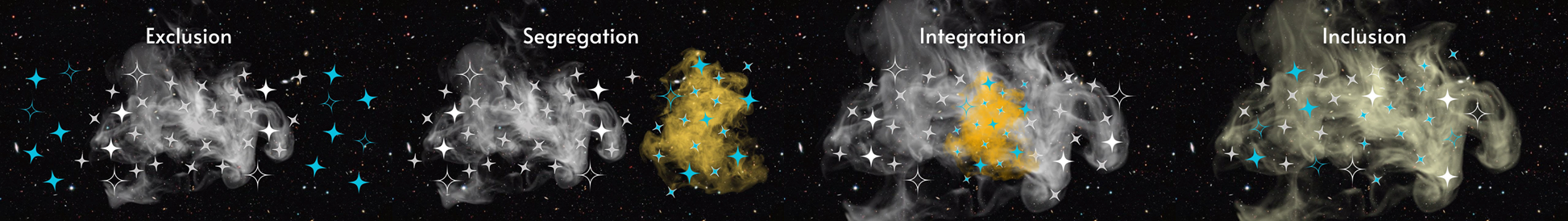}  
    \caption{Several sketches representing the concepts (from left to right) of exclusion, segregation, integration, and inclusion.} 
    \label{inclusion} 
\end{figure}

Inclusion is not a concept easy to understand for those who are not familiar with the teaching or outreach for collectives of people with any disability. In Figure \ref{inclusion}, several basic sketches are used to illustrate the different steps that are necessary to follow before reaching total inclusion in the design of an outreach activity\footnote{To remark the multi-messenger approach of these examples, alternative sketches based on the sense of audition can be explored in \url{https://astroaccesible.iaa.es/content/\%C2\%BFpor-qu\%C3\%A9-astroaccesible-es-un-proyecto-integrador-e-inclusivo\%23:~:text=En\%20Astroaccesible\%2C\%20no\%20s\%C3\%B3lo\%20se,la\%20comunidad\%20cient\%C3\%ADfica\%20y\%20astron\%C3\%B3mica}, as programmed with the tool {\sc Strauss} \cite{strauss}, available at \url{https://github.com/james-trayford/strauss}}. The first necessary step is the recognition of exclusion, including the methods and procedures that can exclude a part of the target public. A first treatment of this exclusion is partially adopted by means of segregation, or the organization of specific and separated actions. A more appropriate strategy can be followed by means of integration, in which people with or without any special need are treated in the same activity, although using different resources. The desirable and ideal scenario is inclusion, in which everyone receives the same information using the same resources. This last is an ideal situation that can be partially reached through a pathway covering the previous stages and adopting an inclusive philosophy in the design of our outreach programs, which enormously helps to improve the outreach to everyone.

Therefore, {\em Astroaccesible} has not just the objective of connecting the BVI colective to science, but also  of convincing other scientists and teaching professionals to incorporate an inclusive aspect to their projects, as this largely benefit the collective of impaired people, helping at the same time to improve the quality of their contents for everyone. This strategy can be considered as a facet of the well-known Universal Design for Learning (UDL) method, already put in practice for other outreach and learning projects related with astronomy and space sciences, like {\em AstroAcess} \cite{astroaccess}.

This not only makes all transmitted concepts easier to understand for everyone but also helps to sensitize the whole population about the importance of inclusion of people with disabilities, letting this collective be more interested in science, and promoting the idea of the incorporation of this collective to begin a professional scientific career, enhancing diversity in research groups, which is demonstrated to increase the productivity and impact factors of the published work \cite{nature2018diversity}.

Among the various activities undertaken by {\em Astroaccesible} in recent years are sessions specifically designed for BVI individuals, often in collaboration with the {\em Organizaci\'on Nacional de Ciegos de Espa\~{n}a} (ONCE). These sessions, however, also cater to a wide range of audiences at different educational levels, from primary school to college, always employing an inclusive methodology. 
In all these activities, it is common to provide as a least inclusive level, not necessarily based on the use of specific resources based on a multisensory approach, comprehensive oral descriptions of the presented material, supplemented by appropriate comparisons, emphasizing the significant role of sight in acquiring the explained information.

\section{A multi-messenger approach}

The outreach and teaching of the new multi-messenger astronomy, involving a wide range of electromagnetic ranges, going from gamma rays down to radio sources, and also involving detection of cosmic rays, neutrinos, or gravitational waves, can be thus done by multiple sensory channels, not just restricted to the use of visual resources, helping the inclusion of BVI. In addition, this strategy also takes advantage of a richness of resources to reinforce the idea of a richness of sources and, as many of these means provide information about the same objects as seen from different energies and perspectives, to provide a more complete view of the different aspects coming from the variety of properties that can be explored.

The adoption of a multi-messenger strategy implies thus the use of a diversity of resources and channels for the transmission of the information not exclusively based on the sense of sight, but extended to other senses. To address this, Astroaccesible has implemented the use of all kind of alternative resources,including  maquettes and models to represent astronomical objects and phenomena, allowing BVI individuals to physically explore the shapes and structures of celestial bodies. Additionally, the project incorporates soundscapes and audio descriptions to convey data and observations in an auditory format. For example, the sonification of data from gravitational waves or the translation of visual graphs into sound patterns can provide an alternative means for BVI individuals to perceive and understand complex astronomical information \cite{PM23}. In the below subsections, more details are given around each one of these adaptations.

\subsection{An inclusive use of images}

One of the most common approaches when dealing with an audience belonging to the BVI colective is a total renunce to the use of images, accepting the prior assumption that all assistants are totally blind. Instead, the preparation of every documention with the Braille language is used. This consideration can be wrong as most BVI individuals  are not totally blind and, in fact, do not dominate the Braille system. On the contrary the preferred system to access towritten  information is much more efficiently done by means of screen readers in mobile phones. Therefore, it is usually a much more accesible system to provide all documentation in an electronic format that can be accesed by every assistant.

For the use of images, there is a similar problem. Apart from the fact that BVI are usually accompanyed by other people without any sight problem that are also participants in the activities, the heterogeneity of visual impairments is not usually considered; most visually impaired people have residual vision or visual memory, so the needs and approaches to make the disseminated information accessible vary greatly (\cite{Benacchio2000}).

\begin{figure}[t]
    \centering
    \includegraphics[width=\textwidth]{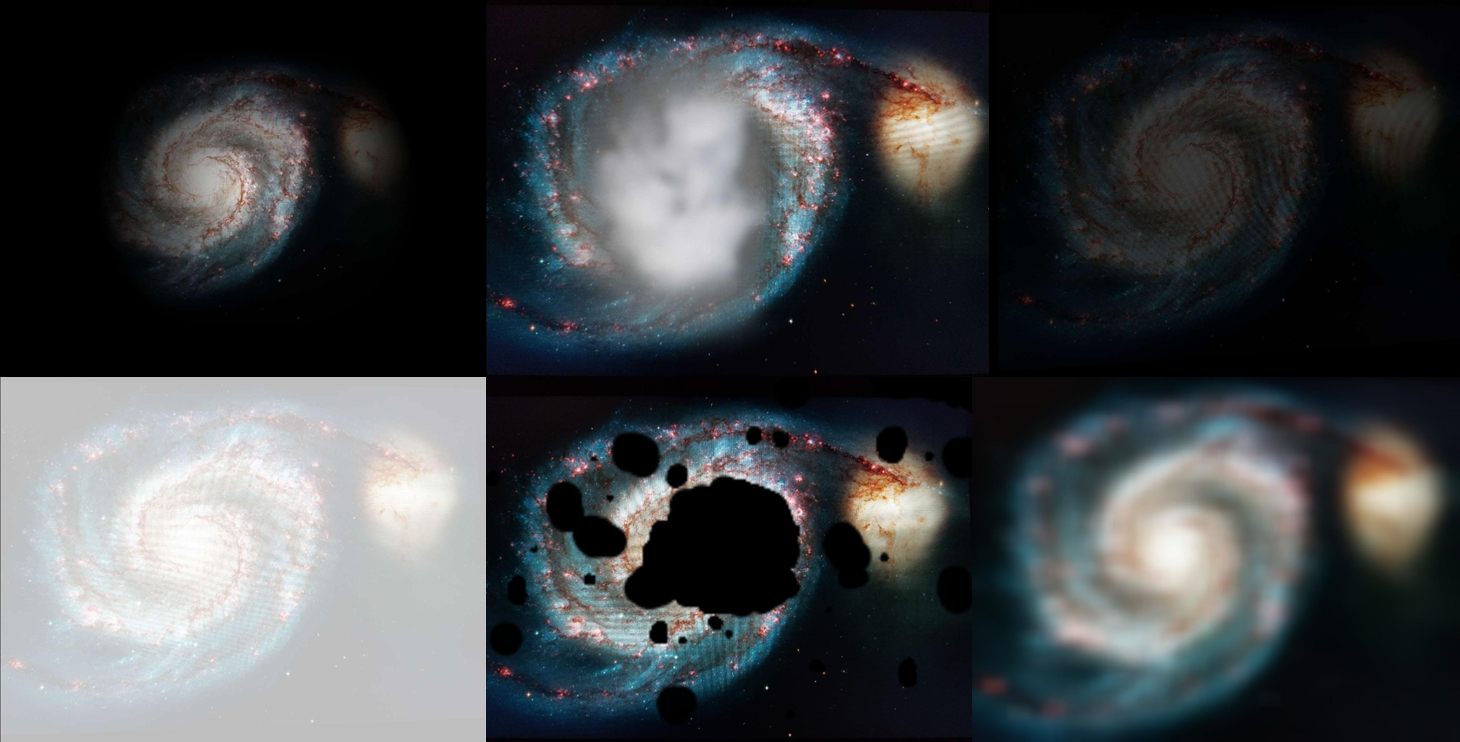}  
    \caption{Distorted images of the spiral galaxy M51 to show how an individual with a visual impairment can perceive it. From left to right and from up to down: a) loss of visual field, b) patched vision, c) loss of central vision, d) night blindness, e) photophobia, and f) loss of visual acuity. These plots were done thanks to the application {\em VR Tengo baja visi{\'o}} developed by the association {\em Begisare}.} 
    \label{low_vision} 
\end{figure}

In Figure \ref{low_vision} it is represented the spiral galaxy M51 as observed through several optical filters, as seen by people with different types of visual affections, including loss of peripherical vision, patched vision, loss of central vision, night blindness, photophobia, or loss of visual acuity. These images were obtained using the application {\sc VR TEngo Baja Visi\'on}, developed by the Spanish association {\em Begisare}. The mobile phone can be also used by any user to try to overcome these difficulties if the images are previously provided in an electronic form. This also includes some solutions based on artificial intelligence able to provide very accurate descriptions. In any case, it is preferable to adopt a preventive strategy trying to show the images under different configurations, contrasts, sizes, and colors to try to let most people  access to the visual information, even if a part of them cannot correctly see it., Thus, one must never renonce to the use of images, given the inclusive strategy taht it is prferable to follow in any activity, giving information in as many different channels as possible.

\begin{figure}[t]
    \centering
    \includegraphics[width=0.53\textwidth]{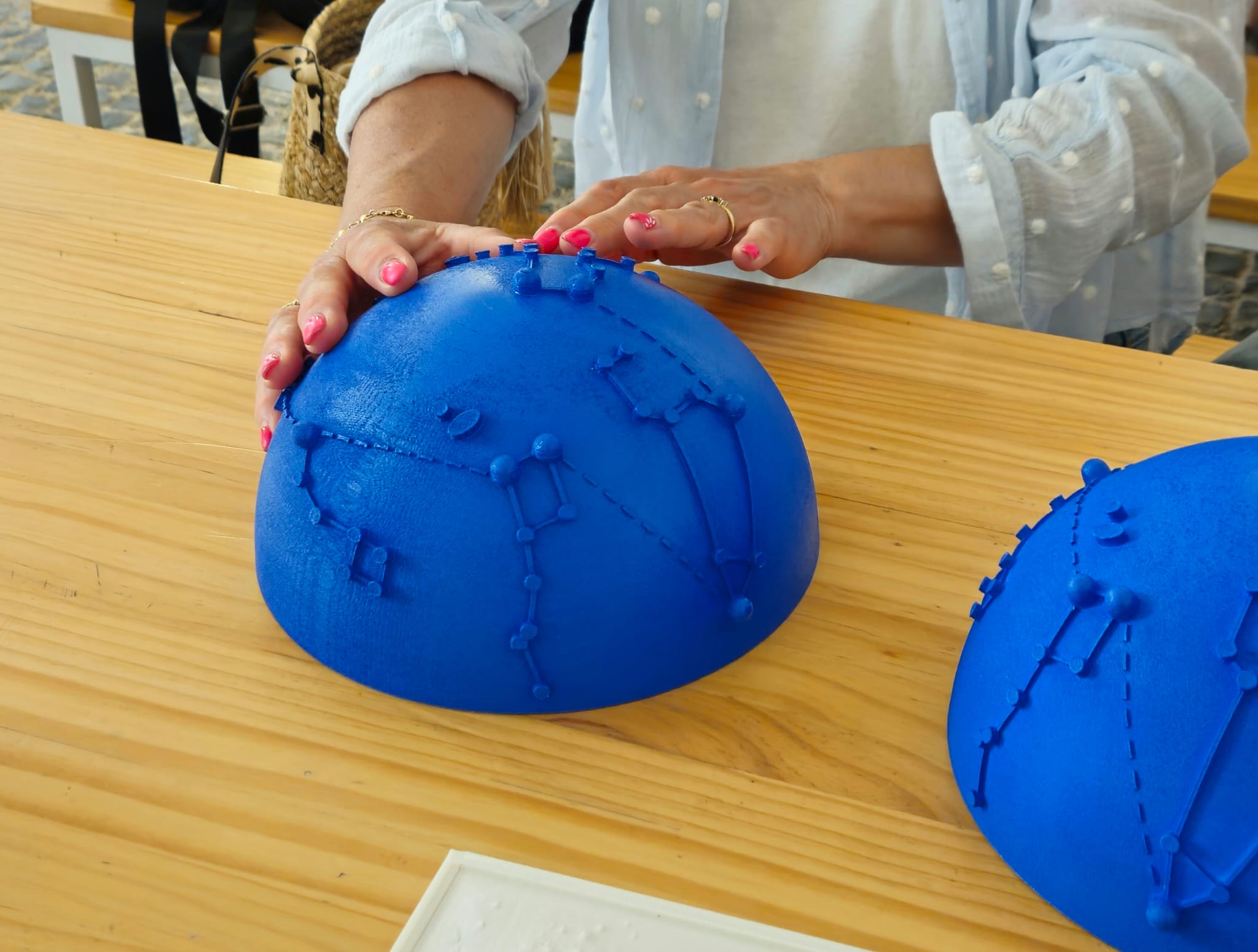}  
\hspace{0.5cm}
    \includegraphics[width=0.38\textwidth]{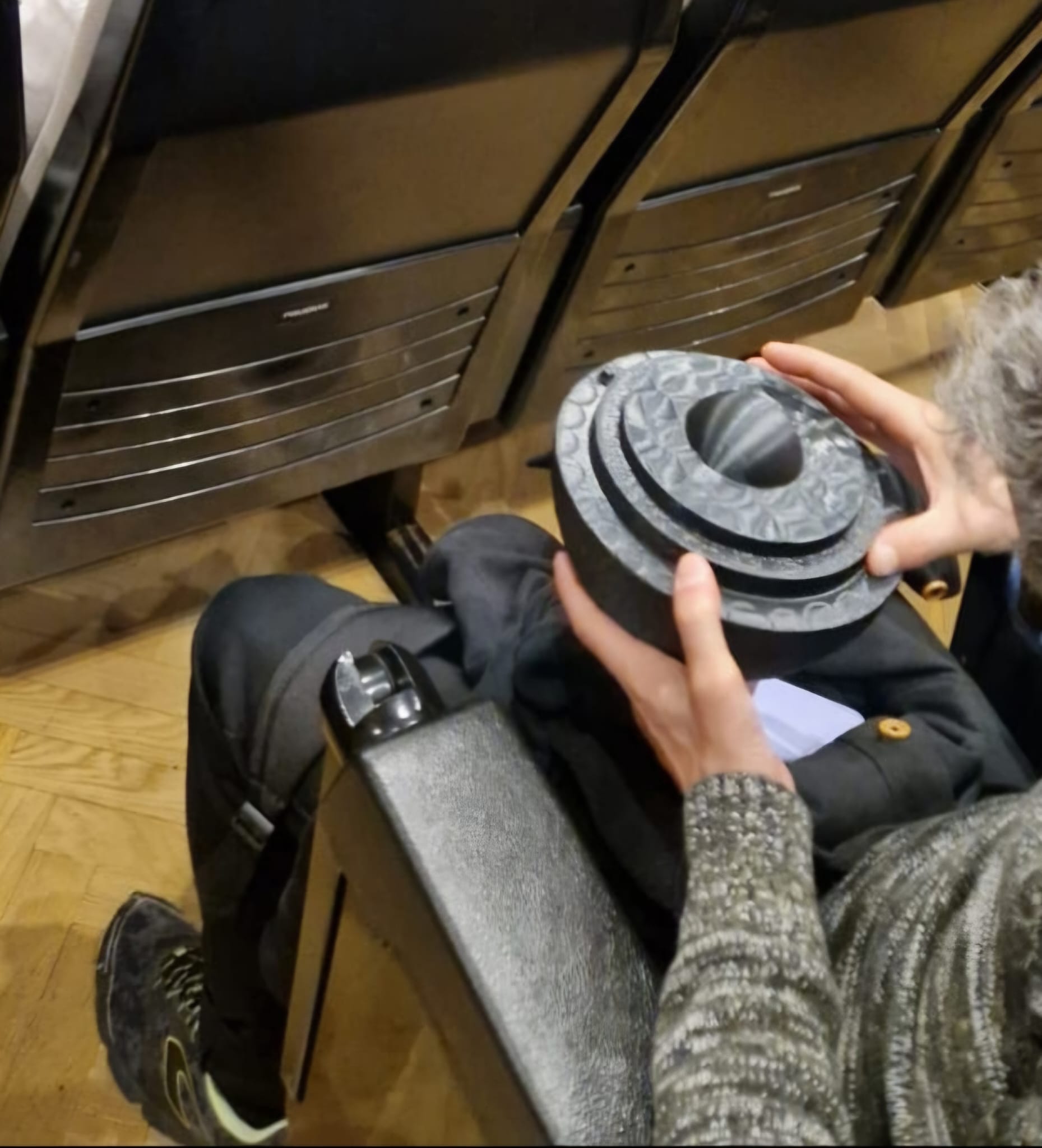}  

    \caption{Two examples of models used in the activities of {\em Astroaccesible}. Left: A model of the night sky as seen from the northern hemisphere. Rigt: A model of the Sun, with its different inner layers.} 
    \label{models} 
\end{figure}

\subsection{Use of tactile material}

Another important resource used in all the activities of {\em Astroaccesible} is the tactile material, both in the form of relief sheets and 3D models. Some of them can be seen in Figure \ref{models}.

Among them, are those developed by the project {\em A Touch of the Universe} \cite{ortiz19}, representing various rocky planets and moons of the Solar System. 
Our models of asteroids and comets, featuring detailed surfaces, help users understand these celestial bodies' shapes and structures.
These are not only useful for a complete recognition of the characteristics of the bodies of the Solar System for BVI, but also for a much more complete identification of their geological features for everyone, overcoming the limitation of a much more simplistic exploration based on a projected image.

Additionally, 3D models of the night sky, printed on the outer surface of a hemisphere and featuring stars in relief connected by constellation lines, have been extensively utilized in our activities. These models assist BVI individuals in forming a spatial mental image of the sky's layout, and they also aid sighted individuals by projecting the same figures onto a screen, facilitating identification and contextual understanding of the night sky. Over the past few years, we have employed 3D models depicting the night sky visible from the Northern Hemisphere during autumn and winter, from Orion to Gemini constellations. These models are particularly useful when public observations are infrequent due to adverse weather conditions.
Moreover, we have developed 3D models representing the summer night sky, including constellations such as Scorpius, Sagittarius, and the Summer Triangle. These models can be used during both outdoor and indoor inclusive activities, providing a comprehensive view of the sky during warmer months. 

 Tactile models of the night sky offer BVI individuals the chance to explore the stars' arrangement and to locate deep space objects, while sighted participants can compare these tactile impressions with visual representations. This approach not only enhances spatial awareness but also enriches the educational experience for all participants. 

There is a pletora of other projects whose tactile resources have been widely used by {\em Astroaccesible}, including the kit {\em Astro TES}, comprising different elements like a model of the Sun\footnote{The project {\em Astro TES: Tocar, Escuchar, SEntir}, leaded by Sebasti\'an Musso, can be accessed via web in \url{https://astrotes.org}}; the {\em Astro BVI} project\footnote{The {\em Astro BVI} project can be visited at \url{https://www.astro4dev.org/category/astrobvi-an-astronomical-educational-kit-for-the-blind-and-vision-impaired-bvi-community-in-south-america/}} , that includes low relief representations of different galaxy types; or a model of a protoplanetary disk\footnote{Developed by the Millenial Planetary Formation group, available  at \url{https://npf.cl/2020/11/10/2213/}}.
All this material, used in combination of images, appropriate descriptions and, as we will see in the next subsection, sonifications, offer a incredible diversity of resources that are not only inclusive, but also expands in a unexpected way the motivation, compromise and  inspiration of every participant for outreach and educational activities.,

\begin{figure}[t]
    \centering
    \includegraphics[width=\textwidth]{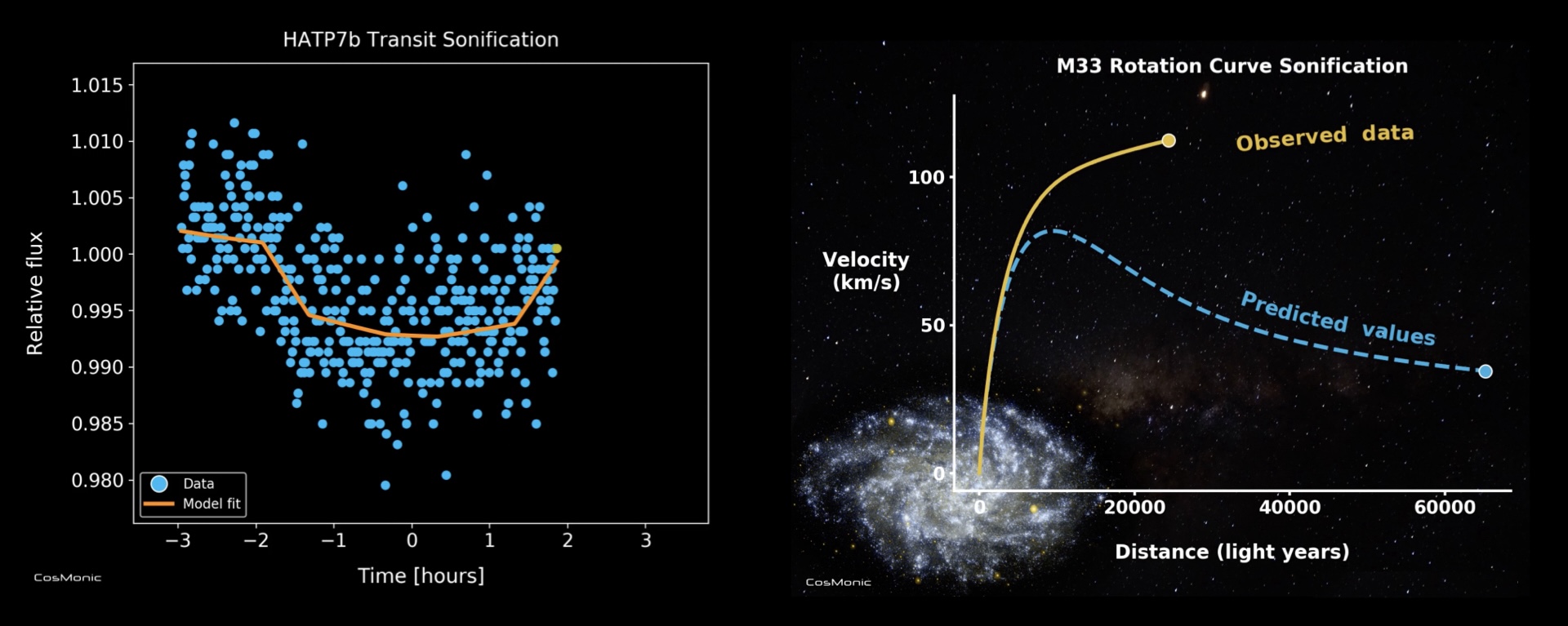}  

    \caption{Two snapshots of animations developed within the project {\em Cosmonic}, sonifying plots. Left: LIght curve of the transit of the exoplanet HATP7. Right: Predicted and observed radial stellar velocity curve around the disk of the spiral disk M33 to explore the effect of the presence of a dark matter halo.} 
    \label{cosmonic} 
\end{figure}

\subsection{Using Sonifications}

Sounds constitute an invaluable resource that can significantly enhance the multisensory aspect of the activities carried out by {\em Astroaccesible}, adding an extra dimension to all presented material. The use of sound, combined with images and tactile models, creates a comprehensive multisensory learning environment. This approach not only facilitates the inclusion of BVI individuals but also enhances the educational experience for all, fostering a deeper understanding of the universe through a multi-messenger perspective \cite{zanella}. There is a growing interest from researchers and sound designers who are increasingly committed to the task of converting astronomical data into sound \cite{rgb23}.

The variety of techniques available for converting astronomical data into sound, such as audification, sonification, and musification, allows for their incorporation into different adapted activities, deepening their inclusive aspect. These techniques enable direct interaction of BVI astronomers with astronomical data, facilitating their integration into research groups in various contexts \cite{Deandra22}. Sonification presents several advantages when used in conjunction with images, helping to present a more complete vision.

For instance, the {\em Cosmonic} project \cite{cosmonic}, whose various products can be accessed on its website \url{http://rgb.iaa.es/es/cosmonic/}, offers different animations with accompanying sounds. These resources enable BVI individuals to access the data, while also helping sighted individuals interpret the graphical content more effectively. The integration of sound not only aids in data comprehension but also enhances the overall sensory experience for all participants. Two examples of the graphics sonified by {\em Cosmonic} are shown in Figure \ref{cosmonic}, including the light curve of the star HATP7 with a transit of an exoplanet, and the radial stellar velocity around the spiral galaxy M33 to illustrate the effect of a dark matter halo.

Beyond the inclusive aspect of sound, explaining the meaning of each property of the sound, including the chosen timbre, loudness, and tone, highlights the arbitrary nature of every sensory channel we use for astronomical data. This feature of sound makes it especially convenient to raise awareness about the role of data processing both for analysis and education. Unlike images, which most people consider the natural way to represent astronomical data, sound can emphasize the conversion factor between a digital signal and its representation, making the process more transparent and understandable.

Additionally, the ability of sound to be perceived at a much higher time resolution by the human ear, as well as the broader frequency range that can be detected, can be particularly useful for specific time-dependent data series \cite{Brown22}. In this context, it is a misconception to associate sounds exclusively with gravitational waves. The quality of sound can represent any type of astronomical data, enriching the multisensory outreach. Sound can convey information about the structure of galaxies, the dynamics of star formation, or the motion of celestial bodies, thus providing a versatile tool for education and inclusion.

\subsection{Audio descriptions: Using language as an additional window}

Given that the basis of a multi-messenger transmission of the astronomical information must reside on the principle of supplying alternative sources to the information, not exclusively based on the use of images to favour inclusion, other sources of information not based on senses can be envisaged. This is the case of language, as a tol to provide information at different levels. In this case, {\em Astroaccesible} tries to foster its use above all in Spanish promoting the diffusion of written texts, easily converted into a spoken speech, covering different matters related with astronomy and science.

In addition, another useful adaptation that {\em Astroaccesible} has implemented is the use of audio description (AD) \cite{EUEP22}.
AD is a valuable access service that translates visual information into words, enhancing the multisensory aspect of activities carried out by {\em Astroaccesible}. This intersemiotic translation practice  bridges visual and linguistic signs, providing a comprehensive understanding of astronomical data through an inclusive, multi-messenger perspective. The primary goal of AD is to compensate for the lack of visual information, ensuring that people with sight loss can comprehend the described source material (e.g., films or paintings) similarly to sighted individuals.

AD practice, traditionally focused on films and audiovisual media, is expanding into new areas such as museums and live events and its potential application in science, particularly astronomy, for both research and educational purposes, is immense. By incorporating AD, {\em Astroaccesible} not only supports BVI individuals but also enhances the educational experience for all participants, promoting a deeper understanding of the universe.

In science communication, AD scripts need to be meticulously crafted to convey essential visual details, aiding BVI individuals in forming mental images of the described objects. This guided viewing experience benefits sighted people as well, enriching their sensory experience and enhancing information retention . The {\sc El Universo en palabras}\footnote{All videos of {\em El Universo en palabras}, can be viewed in {\em YouTube} in \url{https://www.youtube.com/playlist?list=PLDOpkwOM33-YAGnWAKSV9ZHiljg56o8mK}} ({\em The Universe in words}), exemplifies this approach, where final-year Translation and Interpreting students at the University of Granada created audio-described videos under the supervision of both AD trainers and astronomers.

These videos, focusing on very popular astronomical objects, including objects from the Messier catalogue,, provide detailed visual and scientific information, helping both blind and sighted people understand complex astronomical concepts. The project's success underscores the potential of AD as a truly inclusive resource, enhancing the multisensory learning environment by incorporating elements like tactile aids and sonifications.

AD's evolving role in science education highlights its broader application. It helps non-experts and experts alike by focusing attention on critical visual features, thus improving understanding and engagement. This approach is particularly relevant in astronomy, where expert knowledge significantly influences meaning-making processes.

\section{Summary and conclusions}

{\em Astroaccesible} is an astronomy outreach project developed at IAA-CSIC, led by a blind astronomer, which began as an initiative to demonstrate that astronomy can be made accessible to BVI individuals. Over ten years of activities, it has proven that total inclusion is possible by extending strategies and resources originally designed for BVI people to all individuals, regardless of any disability.

Adopting a multi-messenger philosophy for transmitting information that is collected through ground observatories, space telescopes, and underground experiments covering various ranges of electromagnetic radiation or detecting particles—has proven effective. The use of different types of images combined with tactile materials, sounds useful for representing time variations, and audio descriptions that help the general public understand the context of these representations creates an excellent, inclusive set of tools. This approach makes astronomy, astrophysics, and astroparticle physics more accessible and engaging, opening inspiring new ways to handle the diverse ways the Universe sends us information.

Implementing a teaching, outreach, and scientific content dissemination strategy based on Universal Design for Learning (UDL) criteria is essential for including disabled people in the scientific community. This approach encourages them to take an interest in science, potentially leading to professional involvement, and promotes the formation of more diverse research groups capable of finding creative and innovative solutions to scientific problems.

\bibliographystyle{unsrt} 
\bibliography{proceeding}

\end{document}